\documentclass[aps,twocolumn]{revtex4}

\RequirePackage[T2A]{fontenc}

\usepackage{amssymb}
\usepackage{amsmath}
\usepackage{bm}
\usepackage[dvips]{epsfig}
\usepackage{graphicx}

\begin{document}

\title{On the scaling properties of quasicrystalline potentials 
of eightfold rotational symmetry}

\author{A.Ya. Maltsev}

\affiliation{
\centerline{\it L.D. Landau Institute for Theoretical Physics}
\centerline{\it 142432 Chernogolovka, pr. Ak. Semenova 1A,
maltsev@itp.ac.ru}}

\begin{abstract}
 We consider a special class of quasi-periodic potentials arising 
in the physics of photonic systems and possessing rotational 
symmetry of the 8th order. We are interested in the ``scaling'' 
properties of such potentials, namely, the growth rate of their 
closed level lines near the percolation threshold. Estimates of 
the corresponding scaling indices allow, in particular, 
to carry out some comparison of such potentials with various 
models of random potentials on the plane.
\end{abstract}

\maketitle

\section{Introduction}

 In this paper we consider the level lines of two-dimensional 
quasi-periodic potentials with eightfold rotational symmetry, 
which are of considerable interest in the physics of photonic 
systems or systems of ultracold atoms (see, for example, 
\cite{GrynbergRobilliard,ZitoPicSantamatoMarTkachAbbate,
JagannathanDuneau,VShCYuSchn,GautierYaoSanchezPalencia}). 
Potentials of this type are created by standing electromagnetic 
waves and are usually represented by a finite number of Fourier 
harmonics. We will consider here two-dimensional potentials 
$\, V (x, y) \, $, which have the form
\begin{equation}
\label{GenPoten}
V ({\bf r}) \,\,\, = \,\,\, V_{0} \, 
\sum_{j=1}^{4} \, \cos \big( {\bf G}_{j} \cdot {\bf r} 
\, - \, A_{j} \big) \,\,\, , 
\end{equation}
 where $\, {\bf G}_{j} \, $ represent the basis vector 
 $\, \bm{\kappa} \, = \, (k , 0) \, $, rotated by angles 
 $\, 0^{\circ} \, $, $\, 45^{\circ} \, $, $\, 90^{\circ} \, $ 
 and $\, 135^{\circ} \, $ respectively 
 ($A_{j} \in [0, 360^{\circ})$).

 Potentials (\ref{GenPoten}) are quasiperiodic functions 
with 4 quasiperiodes on the plane, i.e. they arise as 
restrictions of a 4-periodic function 
$\, F ({\bf z}) \, = \, F (z^{1}, z^{2}, z^{3}, z^{4} ) \, $ 
in $\, \mathbb{R}^{4} \, $ under some affine embedding 
$\, \mathbb{R}^{2} \rightarrow \mathbb{R}^{4} \, $. 
It is easy to see that in our case we can take
$$F ({\bf z}) \,\,\, = \,\,\, V_{0} \, \big(
\cos z^{1} \, + \, \cos z^{2} \, + \, \cos z^{3} \, + \,
\cos z^{4} \big) \,\,\, , $$
while the embeddings
$\, \mathbb{R}^{2} \rightarrow \mathbb{R}^{4} \, $
are given by the formulas
$${\bf r} \,
\,\, \rightarrow \,\,\, 
\left(
\begin{array}{c}
{\bf G}_{1} \cdot {\bf r} \, - \, A_{1}  \\
{\bf G}_{2} \cdot {\bf r} \, - \, A_{2}  \\
{\bf G}_{3} \cdot {\bf r} \, - \, A_{3}  \\
{\bf G}_{4} \cdot {\bf r} \, - \, A_{4}
\end{array}   
\right)    $$

 In fact, one can see that two of the shift parameters 
$\, {\bf A} \, $ correspond to ``trivial'' shifts of 
the potential $\, V ({\bf r}) \, $ in the plane 
$\, \mathbb{R}^{2} \, $, and it is natural to exclude 
them from consideration. It will be convenient for us 
to introduce here two periodic potentials
\begin{multline*}
V_{1} ({\bf r}) \,\,\, = \,\,\, V_{0} \, \Big(
\cos \big( {\bf G}_{1} \cdot {\bf r} \big) \, + \,
\cos \big( {\bf G}_{3} \cdot {\bf r} \big) \Big) 
\,\,\, =  \\
= \,\,\, V_{0} \, \big( \cos x \, + \, \cos y \big) 
\end{multline*}
and
$$V_{2} ({\bf r}) \,\,\, = \,\,\, V_{0} \, \Big(
\cos \big( {\bf G}_{2} \cdot {\bf r} \big) \, + \,
\cos \big( {\bf G}_{4} \cdot {\bf r} \big) \Big) $$
and consider the potentials
\begin{equation}
\label{VaPot}
V({\bf r}, \, {\bf a}) \,\,\, = \,\,\, 
V_{1} ({\bf r}) \,\, + \,\, V_{2} ({\bf r} - {\bf a}) 
\,\, , \quad \quad
{\bf a} \, = \, (a^{1}, a^{2}) \, \in \, \mathbb{R}^{2} 
\end{equation}

 Both potentials $\, V_{1} ({\bf r}) \, $ and 
$\, V_{2} ({\bf r}) \, $ have rotational symmetry 
of order 4, it is easy to see also that the potential 
$\, V_{2} ({\bf r}) \, $ represents the potential 
$\, V_{1} ({\bf r}) \, $ rotated by $45^{\circ}$ 
relative to the origin.

 The potential
$$V({\bf r}, \, {\bf 0}) \,\,\, = \,\,\, 
V_{1} ({\bf r}) \,\, + \,\, V_{2} ({\bf r}) $$
has exact rotational symmetry of the 8th order and is 
quasiperiodic. It is this potential that is, in fact, 
considered most often in the physics of two-dimensional 
systems. Here, however, it will be convenient for us 
to consider the entire family of quasiperiodic 
potentials (\ref{VaPot}).

 Here we will be interested in the geometry of the level 
lines of potentials $\, V({\bf r}, \, {\bf a}) \, $:
\begin{equation}
\label{Vepsilon}
V({\bf r}, \, {\bf a}) \,\,\, = \,\,\, \epsilon \,\,\, ,
\end{equation}
as well as the geometry of the areas
$$V({\bf r}, \, {\bf a}) \,\,\, < \,\,\, \epsilon
\quad \quad \text{and} \quad \quad 
V({\bf r}, \, {\bf a}) \,\,\, > \,\,\, \epsilon $$
bounded by them. 

 The description of the level lines of quasiperiodic 
functions on the plane is the content of the problem 
of S.P. Novikov, which has been studied quite deeply 
by now (see, for example, 
\cite{MultValAnMorseTheory,zorich1,dynn1992,Tsarev,dynn1,
zorich2,DynnBuDA,dynn2,dynn3,NovKvazFunc,DynNov}). 
We also note here that the case of 4 quasiperiods was most 
deeply studied in the works \cite{NovKvazFunc,DynNov}.

 The basis for considering S.P. Novikov's problem is 
the study of open (unclosed) level lines (\ref{Vepsilon}), 
which largely determine the overall picture in 
$\, \mathbb{R}^{2} \, $. It can be immediately noted 
that open level lines (\ref{Vepsilon}) may appear 
in an energy range narrower than the full range of values 
$\, [ \epsilon_{\min} , \, \epsilon_{\max} ] \, $
of the potential $\, V({\bf r}) \, $.

 A feature of quasiperiodic potentials is that they, 
in a sense, occupy an intermediate position between 
periodic and random potentials. This is particularly 
evident in the fact that many quasiperiodic potentials 
have ``topologically regular'' open level lines.
 
\begin{figure}[t]
\begin{center}
\includegraphics[width=\linewidth]{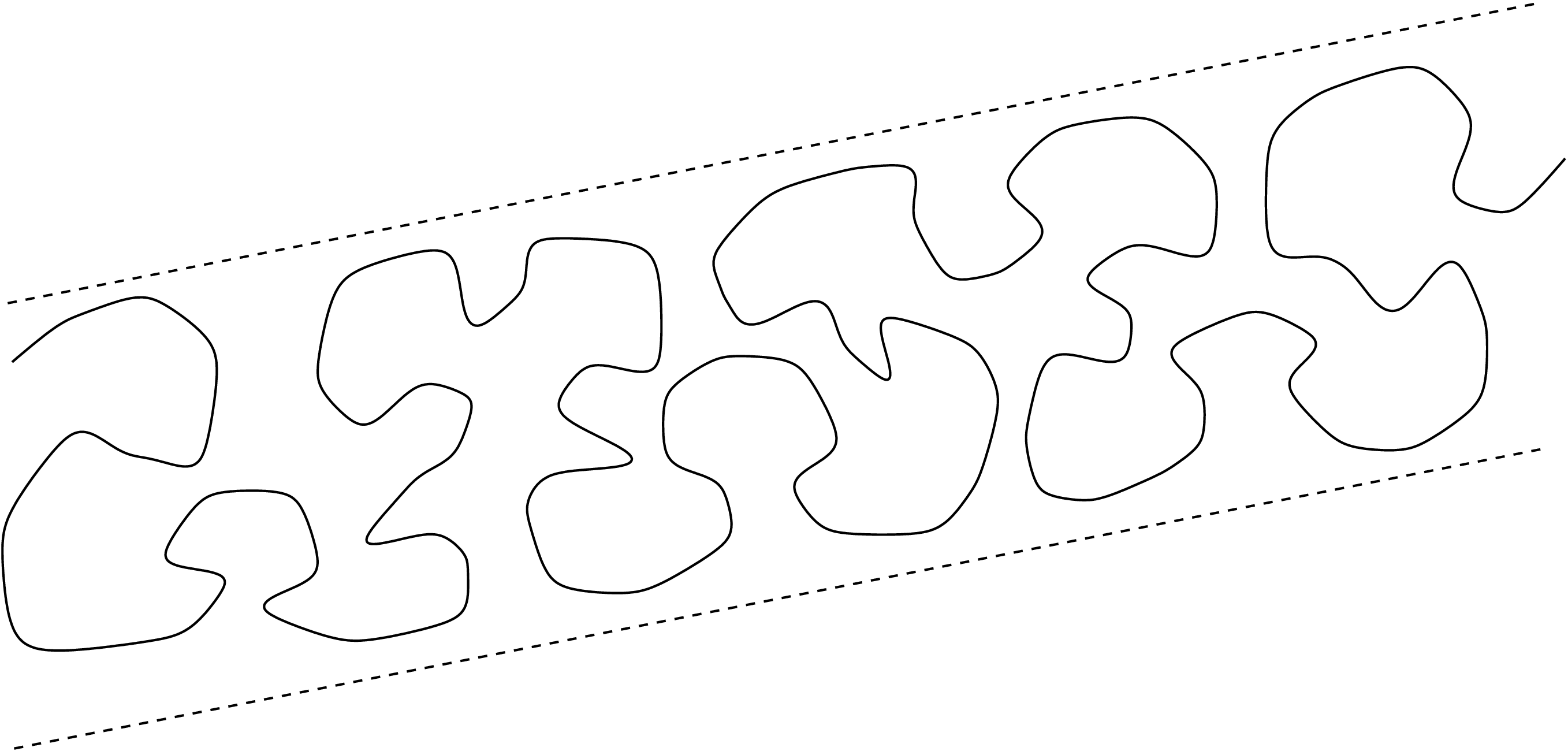}
\end{center}
\caption{``Topologically regular'' level line of a 
quasiperiodic potential (schematically)}
\label{TopRegular}
\end{figure}

 Topologically regular open level lines (\ref{Vepsilon}) 
are not periodic, however, each such level line lies 
in a straight strip of finite width, passing through it 
(Fig. \ref{TopRegular}). In addition, topologically 
regular lines (\ref{Vepsilon}) are stable with respect 
to small variations of the potential parameters and 
usually arise in some finite energy interval 
$\, \epsilon \, \in \, [ \epsilon_{1} , \, \epsilon_{2} ] \, $.
 It can be seen, therefore, that potentials $\, V (x, y) \, $, 
 possessing topologically regular level lines, are in a sense 
 closer to periodic potentials than to random ones. It should 
 be noted that the occurrence of topologically regular level 
 lines plays a very important role in considering a number 
 of questions, and potentials possessing such level lines
 often correspond to rather rich sets in the space 
 of the problem parameters (see, for example, 
 \cite{zorich1,dynn3,NovKvazFunc,DynNov,PismaZhETF,UFN,
 BullBrazMathSoc,UMNObzor,DynMalNovUMN,AnnPhys}). 
 In particular, the occurrence of such potentials is also 
 typical for many important families of potentials 
 with 4 quasiperiods (see \cite{NovKvazFunc,DynNov,AnnPhys}).

 Potentials (\ref{VaPot}), however, cannot have topologically 
regular open level lines, which is due to the presence of 
rotational symmetry in the potentials $\, V_{1} \, $ and 
$\, V_{2} \, $. As was also shown in \cite{Superpos}, open 
level lines (\ref{Vepsilon}) can arise in this case only at 
a single level $\, \epsilon_{0} \, $ 
(for any value of $\, {\bf a}$), which brings such potentials 
closer to random potentials on a plane.

\begin{figure}[t]
\begin{center}
\includegraphics[width=\linewidth]{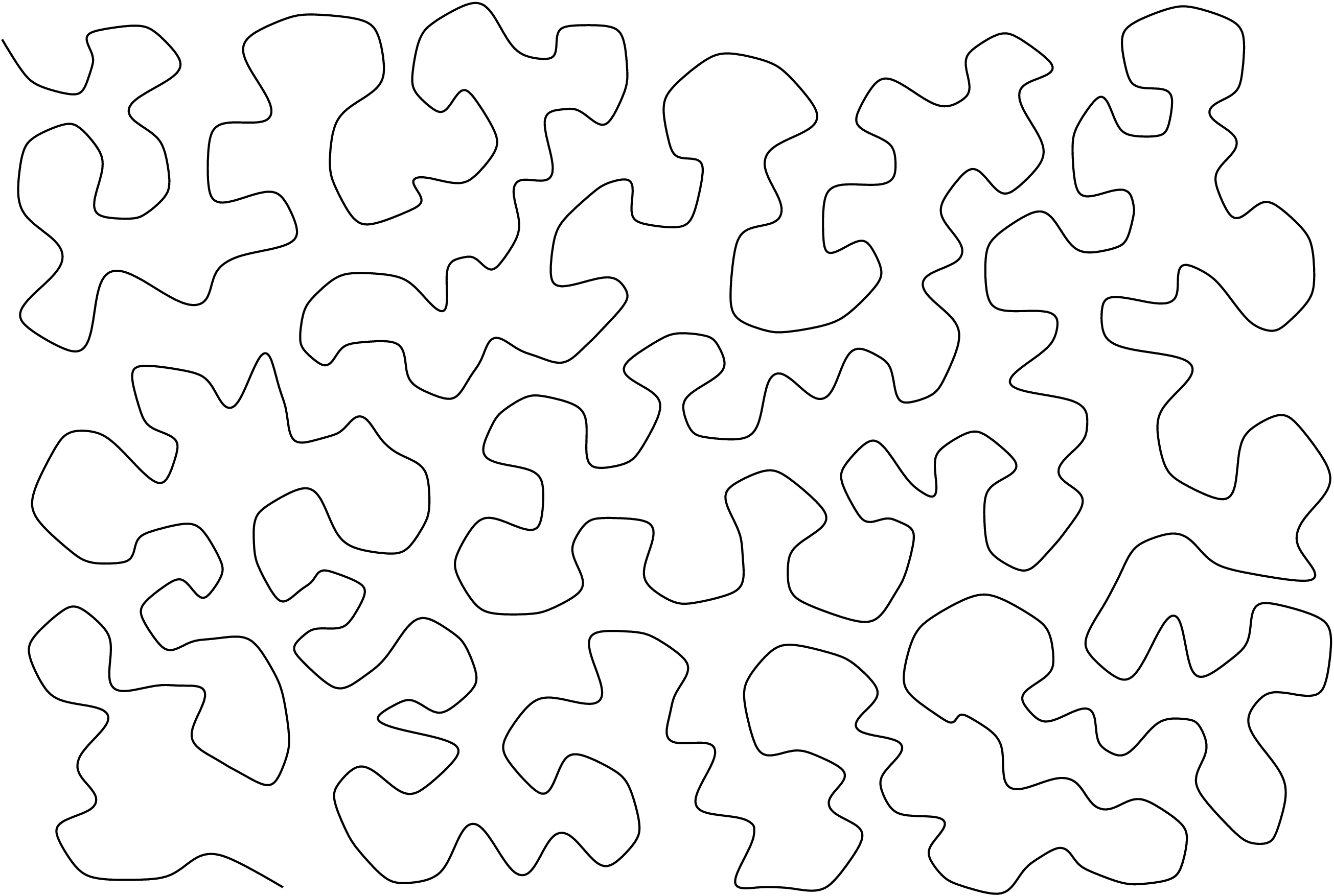}
\end{center}
\caption{``Chaotic'' level line of a quasiperiodic 
potential (schematically)
}
\label{Chaotic}
\end{figure}

 Open level lines of quasiperiodic potentials that are not 
topologically regular have a more complex geometry, wandering 
around the plane in a rather complex manner (Fig. \ref{Chaotic}). 
We will call such level lines ``chaotic''. Many aspects of the 
geometry of chaotic level lines for potentials 
with 3 quasiperiods were studied in the works
\cite{Tsarev,DynnBuDA,dynn2,Zorich1996,ZorichAMS1997,Zorich1997,
zorich3,DeLeo1,DeLeo2,DeLeo3,ZorichLesHouches,DeLeoDynnikov1,dynn4,
DeLeoDynnikov2,Skripchenko1,Skripchenko2,DynnSkrip1,DynnSkrip2,
AvilaHubSkrip1,AvilaHubSkrip2,TrMian,DynHubSkrip}. 
In particular, their behavior often has ``scaling'' properties. 
Here it can be noted that potentials with 4 quasi-periods 
can have chaotic level lines of even more complex geometry.

 Let us note one more circumstance here. As we have already 
said, it will be convenient for us to consider the family of 
potentials (\ref{VaPot}) as a whole, for all values 
of $\, {\bf a} \, $ at once. The value $\, \epsilon_{0} \, $
corresponds in fact to the emergence of open level 
lines of $\, V ({\bf r} , \, {\bf a}) \, $ at least for some 
values of $\, {\bf a} \, $ (possibly not for all). According
to \cite{DynMalNovUMN,BigQuas}, however, in the case when 
open level lines arise (at $\, \epsilon = \epsilon_{0} \, $) 
not for all values of $\, {\bf a} \, $, all potentials 
$\, V ({\bf r} , \, {\bf a}) \, $ must contain closed level lines 
of arbitrarily large size at the level $\, \epsilon_{0} \, $. 
It can be seen, therefore, that the maximal radius of the level 
lines $\, V ({\bf r} , \, {\bf a}) \, = \, \epsilon_{0} \, $, 
as well as of the regions
$\, V ({\bf r} , \, {\bf a}) \, < \, \epsilon_{0} \, $ and 
$\, V ({\bf r} , \, {\bf a}) \, > \, \epsilon_{0} \, $ 
in any case tends to infinity. The above remark allows, 
in particular, to consider all potentials 
$\, V ({\bf r} , \, {\bf a}) \, $ from one point of view 
when considering finite-size systems. We also note that 
the values $\, {\bf a} \, $ corresponding to the emergence
of open level lines 
$\, V ({\bf r} , \, {\bf a}) \, = \, \epsilon_{0} \, $, 
in any case form an everywhere dense set in the space 
of all values $\, {\bf a} \, $.

 It can be seen that the described behavior of the level 
lines of $\, V ({\bf r} , \, {\bf a}) \, $ make these 
potentials similar to random potentials on the plane. 
In general, families of quasi-periodic potentials with 
such properties can be considered as some models 
of random potentials with long-range order.

 For the family $\, V ({\bf r} , \, {\bf a}) \, $ it is 
also easy to show that $\, \epsilon_{0} = 0 \, $. Indeed, 
returning to the representation (\ref{GenPoten}), 
one can see that the transformation
$$\left( A^{1} ,  A^{2} ,  A^{3} ,  A^{4} \right) 
\,\,\, \rightarrow \,\,\, 
\left( A^{1} + \pi , \, A^{2} + \pi  , \, A^{3} + \pi , \, 
A^{4} + \pi \right) $$
corresponds to the replacement
$$V ({\bf r} , \, {\bf a}) \,\,\, \rightarrow \,\,\,  - \,
V ({\bf r} , \, {\bf a}) $$

 Thus, if open level lines arise in the family 
$\, V ({\bf r} , \, {\bf a}) \, $ at 
$\, \epsilon = \epsilon_{0} \, $ 
(at least for some $\, {\bf a}$) they must also arise 
at $\, \epsilon \, = \, - \, \epsilon_{0} \, $. Assuming, 
however (according to \cite{Superpos}), that the value 
$\, \epsilon_{0} \, $ is unique for the whole family 
$\, V ({\bf r} , \, {\bf a}) \, $, we immediately obtain 
then $\, \epsilon_{0} = 0 \, $.

 For $\, \epsilon \neq 0 \, $ all the level lines 
(\ref{Vepsilon}) are closed. Their sizes at each 
level $\, \epsilon \, $ are limited by one constant 
$\, D (\epsilon) \, $, which tends to infinity as 
$\, \epsilon \rightarrow 0 \, $. It can also be seen 
that the closed level lines (\ref{Vepsilon}) 
have a rather simple shape and small sizes near the
values $\, \epsilon_{\min} \, $ and $\, \epsilon_{\max} \, $
and can become significantly more complicated as 
$\, \epsilon \, $ approaches $\, \epsilon_{0} \, $. 
The sizes of the level lines (\ref{Vepsilon}) for 
quasiperiodic potentials usually grow in this case 
according to some power law corresponding to 
certain ``scaling'' properties of the potential.

 In this paper we will be interested in the behavior 
of the constant $\, D (\epsilon) \, $ in the limit 
$\, \epsilon \rightarrow 0 \, $. In particular, we obtain 
here the estimate
$$D (\epsilon) \quad \leq \quad  {\rm const} \,\,  \cdot  
| \epsilon |^{-1} $$
for the potentials under consideration. It can be seen 
that this estimate, generally speaking, differs from 
the known estimates in the theory of percolation and 
random potentials (see, for example,
\cite{Stauffer, Essam, Riedel, Trugman}), 
which again indicates some difference between 
quasiperiodic and random potentials on the plane.
In \cite{TwoLayer} a similar result was obtained 
for a superposition of potentials with rotational 
symmetry of order 3. Here we consider in detail 
the symmetries of order 4 and 8, and also obtain 
more precise estimates using the explicit form 
of the potentials $\, V ({\bf r} , \, {\bf a}) \, $.

 In Section 2 we describe a broader family of 
quasi-periodic potentials that we need to study 
the properties of the potentials given above. 
In Section 3 we present the necessary calculations 
that allow us to obtain estimates of the ``scaling'' 
parameters for the potentials 
$\, V ({\bf r} , \, {\bf a}) \, $.

\section{Extended family of quasiperiodic potentials}
\setcounter{equation}{0}

 As we have already said, the potential 
$\, V ({\bf r} , \, {\bf 0}) \, $, which has exact 
rotational symmetry of the 8th order, is usually of 
greatest interest. To study it, however, we need 
here a wider class of potentials
\begin{multline*}
V ({\bf r} , \, \alpha , \, {\bf a}) \,\,\, = \,\,\, 
V_{1} ({\bf r}) \,\, + \,\, 
V_{2}^{\alpha} ({\bf r} , \, {\bf a}) \,\,\, \equiv  \\
\equiv  \,\,\, 
V_{1} ({\bf r}) \,\, + \,\, 
V_{1} \big( \pi_{-\alpha} [{\bf r} - {\bf a}] \big) \,\,\, , 
\end{multline*}
given by superpositions of the potential 
$\, V_{1} ({\bf r}) \, $ and its
rotation by an angle $\, \alpha \, $ relative 
to the origin and a shift by a vector $\, {\bf a} \, $ 
in the plane $\, \mathbb{R}^{2} \, $ (here we will denote 
by $\, \pi_{\alpha} \left[ {\cal F} \right] \, $ the 
rotation of any figure $\, {\cal F} \, $ in 
$\, \mathbb{R}^{2} \, $ by an angle $\, \alpha \, $ 
relative to the origin).

 As is easy to see, all potentials 
$\, V ({\bf r} , \, \alpha , \, {\bf 0}) \, $ have 
exact rotational symmetry of order 4 and are quasiperiodic 
for generic angles $\, \alpha \, $. In general, 
potentials $\, V ({\bf r} , \, \alpha , \, {\bf a}) \, $ 
do not have exact rotational symmetry. As follows from 
the results of \cite{Superpos}, for generic (not ``magic'')
angles $\, \alpha \, $, open level lines of potentials 
$\, V ({\bf r} , \, \alpha , \, {\bf a}) \, $ may also 
arise only at a single energy value $\, \epsilon_{0} \, $ 
(in this case $\, \epsilon_{0} = 0$).

\begin{figure}[t]
\begin{center}
\includegraphics[width=\linewidth]{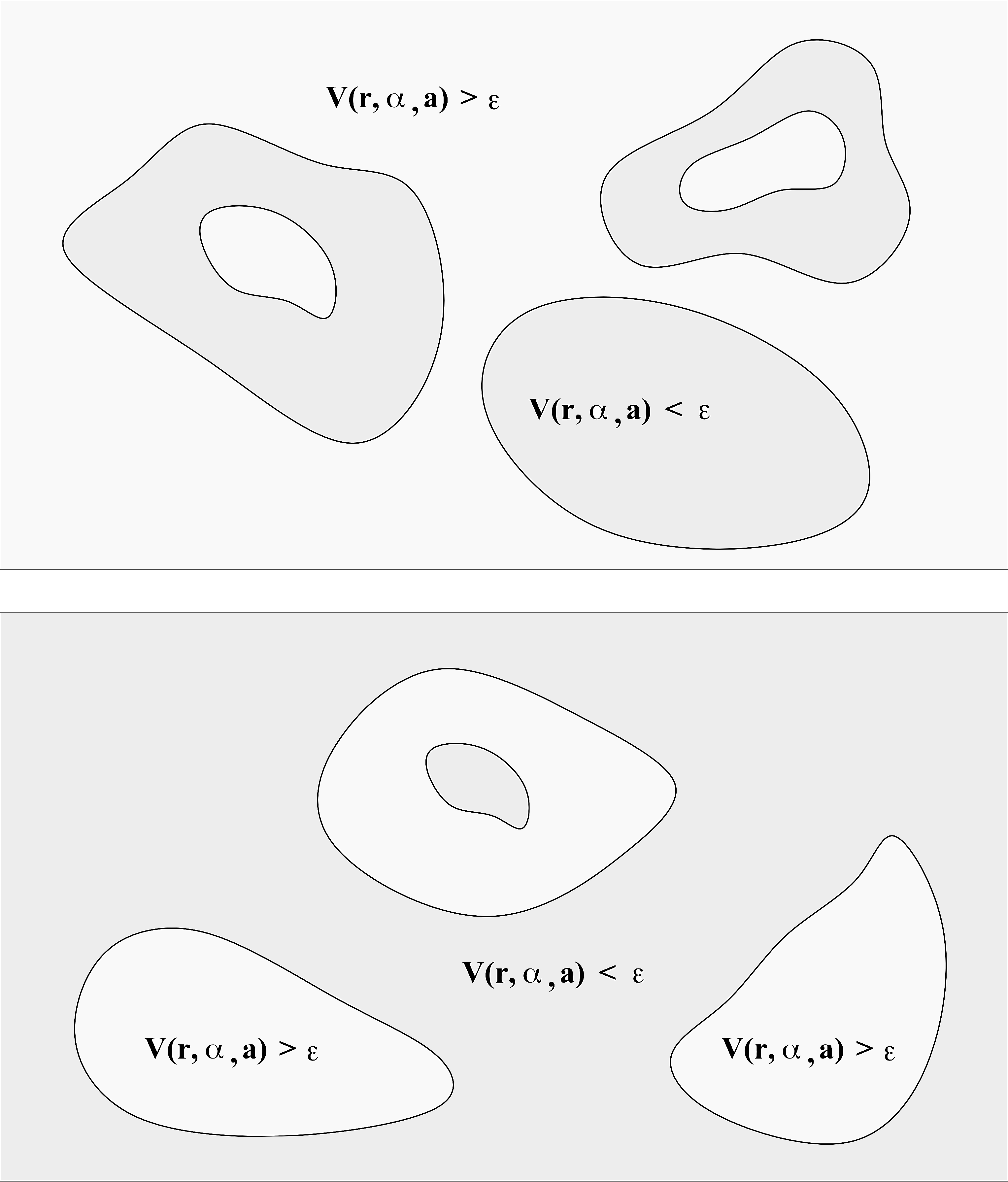}
\end{center}
\caption{Situations of type $\, A_{-} \, $ and 
$\, A_{+} \, $ in the plane $\, \mathbb{R}^{2} \, $ 
(schematically)}
\label{Aminusplus}
\end{figure}

 For generic angles $\, \alpha \, $ and any values of 
$\, {\bf a} \, $ the values $\, \epsilon < 0 \, $
correspond to a situation ``of type $\, A_{-}$'' for 
the corresponding potentials 
$\, V ({\bf r} , \, \alpha , \, {\bf a}) \, $. 
Namely, the set
$$V ({\bf r} , \, \alpha , \, {\bf a}) \,\,\, > \,\,\, 
\epsilon $$
has in this case a unique unbounded component in the plane 
$\, \mathbb{R}^{2} \, $, while all other connected 
components of the sets 
$\, V ({\bf r} , \, \alpha , \, {\bf a}) > \epsilon \, $ 
and 
$\, V ({\bf r} , \, \alpha , \, {\bf a}) < \epsilon \, $ 
are bounded (Fig. \ref{Aminusplus}).

 Similarly, for $\, \epsilon > 0 \, $ the quasiperiodic 
potentials $\, V ({\bf r} , \, \alpha , \, {\bf a}) \, $ 
correspond to a situation ``of type $\, A_{+}$''. 
Namely, the set
$$V ({\bf r} , \, \alpha , \, {\bf a}) \,\,\, < \,\,\, 
\epsilon $$
has in this case a unique unbounded component in the 
plane $\, \mathbb{R}^{2} \, $, while all other connected 
components of the sets 
$\, V ({\bf r} , \, \alpha , \, {\bf a}) < \epsilon \, $ 
and 
$\, V ({\bf r} , \, \alpha , \, {\bf a}) > \epsilon \, $ 
are bounded (Fig. \ref{Aminusplus}).

\begin{figure}[t]
\begin{center}
\includegraphics[width=\linewidth]{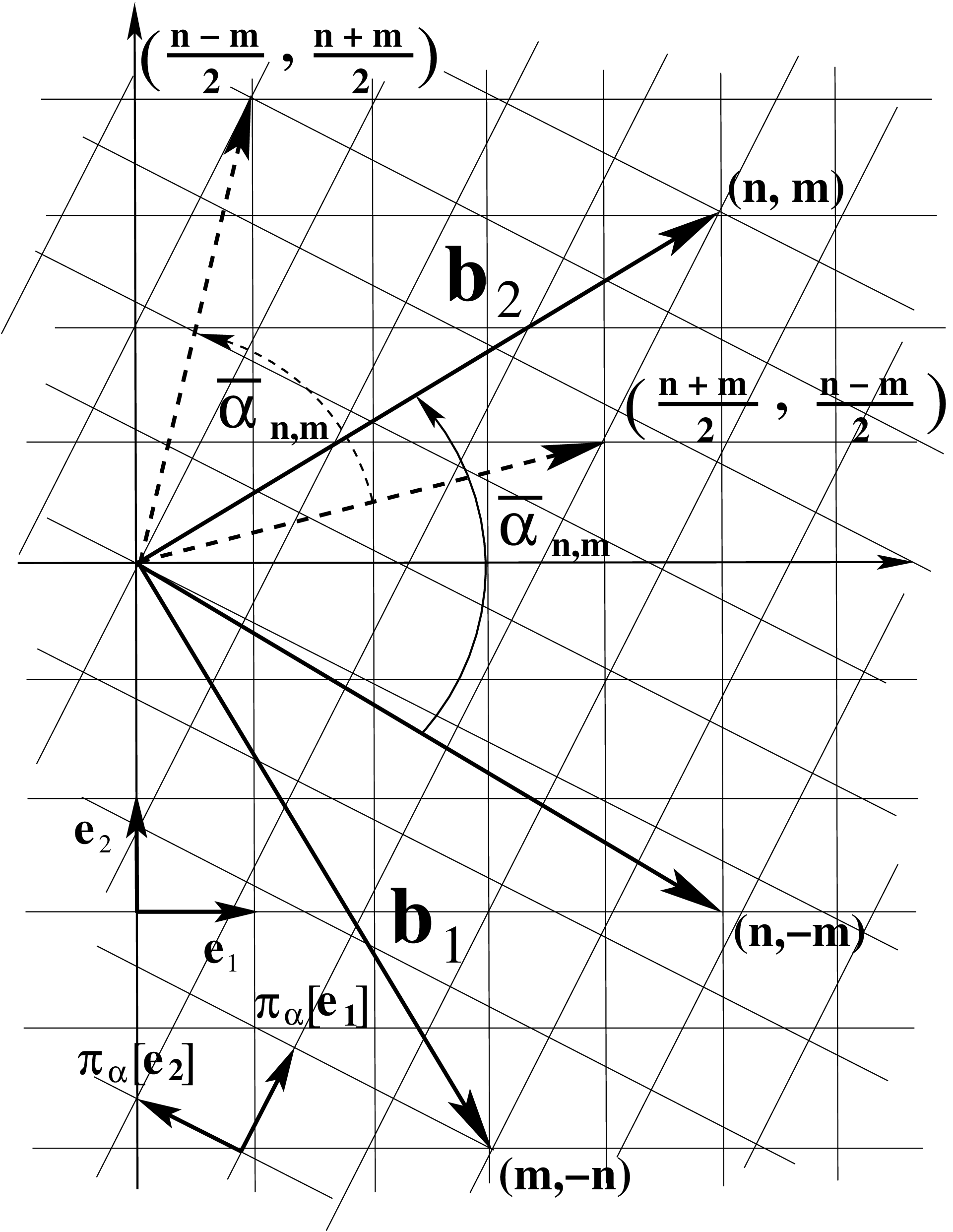}
\end{center}
\caption{Rotation from the vector $\, (n, \, - m) \, $ 
to the vector $\, (n, \, m) \, $, and also from the 
vector $\, (1/2) (n + m , \, n - m ) \, $ to the vector 
$\, (1/2) (n - m , \, n + m ) \, $ in the integer lattice.}
\label{nmPovorot}
\end{figure}

 As is well known, for special (``magic'') angles 
$\, \alpha \, $ the potentials 
$\, V ({\bf r} , \, \alpha , \, {\bf a}) \, $ 
are periodic (with large periods). As is also well 
known, ``magic'' angles $\, \alpha \, $ are associated 
with integer ``Pythagorean'' triples. Here we will use 
the simplest description of ``magic'' angles 
$\, \bar{\alpha}_{n,m} \, $ corresponding to periodic 
potentials $\, V ({\bf r} , \, \alpha , \, {\bf a}) \, $. 
Namely, we define the ``magic'' angle 
$\, \bar{\alpha}_{n,m} \, $ as the rotation angle from 
the vector $\, (n, \, - m) \, $ to the vector 
$\, (n, \, m) \, $ in the standard integer lattice 
(Fig. \ref{nmPovorot}).

 It is easy to see that, due to the symmetry of the 
potential $\, V_{1} ({\bf r}) \, $, we can restrict ourselves 
to the interval $\, \alpha \in (0, 90^{\circ}) \, $ and thus set
$$0 \,\,\, < \,\,\, m \,\,\, < \,\,\, n $$
for the ``magic'' angles $\, \bar{\alpha}_{n,m} \, $. It is 
also easy to see that the periods of the potentials 
$\, V ({\bf r} , \, \bar{\alpha}_{n,m} , \, {\bf a}) \, $ 
are the vectors
\begin{equation}
\label{bPeriods}
{\bf b}_{1} \,\,\, = \,\,\, {2 \pi \over k} \left(
\begin{array}{c}
m  \\
- n 
\end{array}  \right) \quad  \text{and} \quad \quad
{\bf b}_{2} \,\,\, = \,\,\, {2 \pi \over k} \left(
\begin{array}{c}
n  \\
m 
\end{array}  \right)
\end{equation}
(Fig. \ref{nmPovorot}). 

 The numbers $\, m \, $ and $\, n \, $ are assumed 
to be relatively prime, and the vectors (\ref{bPeriods}) 
are the smallest periods of the potentials
$\, V ({\bf r} , \, \bar{\alpha}_{n,m} , \, {\bf a}) \, $ 
if $\, m \, $ and $\, n \, $ have different parities. 
If $\, m \, $ and $\, n \, $ are both odd, the smallest 
periods of the potentials 
$\, V ({\bf r} , \, \bar{\alpha}_{n,m} , \, {\bf a}) \, $ 
are the vectors
$${\bf b}^{\prime}_{1} \,\,\, = \,\,\, {1 \over 2} \left(
{\bf b}_{1} \, + \, {\bf b}_{2} \right) 
\quad  \text{and} \quad \quad
{\bf b}^{\prime}_{2} \,\,\, = \,\,\, {1 \over 2} \left(
- {\bf b}_{1} \, + \, {\bf b}_{2} \right)
\,\,\, , $$
 having length
 $${2 \pi \over k} \sqrt{{m^{2} + n^{2} \over 2}} $$
 
  In the latter case, as is easy to verify, the angle 
$\, \bar{\alpha}_{n,m} \, $ is also the angle of rotation 
from the integer vector
$$\left( {n + m \over 2} , \, {n - m \over 2} \right) 
\quad \text{to the vector} \quad
\left( {n - m \over 2} , \, {n + m \over 2} \right) $$
(Fig. \ref{nmPovorot}). 

 Assuming here
$$ (m_{0}, n_{0}) \,\,\, = \,\,\, (m, n) $$
if $\, m\, $ and $\, n\, $ have different parities, and
$$(m_{0}, n_{0}) \,\,\, = \,\,\, \left( {m + n \over 2} , \,
{m - n \over 2} \right) $$
if $\, m\, $ and $\, n\, $ are both odd, we can
use the quantity
$${2 \pi \over k} \sqrt{m_{0}^{2} + n_{0}^{2}} $$
as the length of the minimal periods of the potentials 
$\, V ({\bf r} , \, \bar{\alpha}_{n,m} , \, {\bf a}) \, $.

 To study the properties of the potential 
 $\, V ({\bf r} , {\bf 0}) \, = \,
V ({\bf r} , \, 45^{\circ} , \, {\bf 0}) \, $ 
(as well as all
$\, V ({\bf r} , \, \alpha , \, {\bf a}) $) we will use 
here their approximations by the potentials 
$\, V ({\bf r} , \, \bar{\alpha}_{n,m} , \, {\bf a}) \, $. 
Approximation of quasiperiodic potentials by periodic ones is, 
certainly, possible only in a bounded region of  
$\, \mathbb{R}^{2} \, $, and we need to use approximations 
of angles $\, \alpha \, $ by the ``magic'' angles 
$\, \bar{\alpha}_{n,m} \, $.

 Obviously, in the general case such approximations are 
directly related to approximations of the quantity 
$\, \tan \alpha / 2 \, $ by rational fractions $\, m / n \, $. 
Taking into account the relation
$$\left| \arctan^{\prime} x \right| \,\,\, \leq \,\,\, 1 
\,\,\, , $$
it is easy to see that the relation
$$\left| \tan {\alpha \over 2} \, - \, {m \over n} \right|
\,\,\, < \,\,\, \delta $$
always implies
$$\left| \alpha \, - \, 2 \arctan  {m \over n} \right| 
\,\,\, \equiv \,\,\, 
\left| \alpha \, - \, \bar{\alpha}_{n,m} \right| 
\,\,\, < \,\,\, 2 \delta $$

 For good approximations of the angle $\, \alpha \, $ by 
the angles $\, \bar{\alpha}_{n,m} \, $ and 
$$\left| \tan {\alpha \over 2} \, - \, {m \over n} \right|
\,\,\, = \,\,\, \delta $$
we can also assume with good accuracy
\begin{multline}
\label{tanSootn}
\left| \alpha \, - \, \bar{\alpha}_{n,m} \right| 
\,\,\, \simeq \,\,\, 
{2 \delta \over 1 + \tan^{2} \alpha / 2 } 
\,\,\, =  \,\,\, 2 \delta \, \cos^{2} {\alpha \over 2} 
\,\,\, =  \\
= \,\,\, \delta \, \big( 1 + \cos \alpha \big) 
\end{multline}

 Potentials $\, V ({\bf r} , \, \alpha , \, {\bf a}) \, $ 
satisfy the relations
$$ V \big( 
{\bf r}, \, \alpha , \,\, {\bf a} + \pi_{\alpha} 
\left[ {\bf e}_{1,2} \right] \big)
\,\,\, \equiv \,\,\, V ({\bf r}, \, \alpha, \, {\bf a} ) 
\,\,\, , $$
$$V \big( 
{\bf r}, \, \alpha , \,\, {\bf a} \, + \, {\bf e}_{1,2} \big)
\,\,\, \equiv \,\,\, V \big( {\bf r} \, - \, {\bf e}_{1,2}, 
\,\, \alpha, \, {\bf a} \big) \,\,\, , $$
where 
$${\bf e}_{1} \,\,\, = \,\,\, (T, \, 0) \,\,\, , \quad 
{\bf e}_{2} \,\,\, = \,\,\, (0, \, T) \quad \quad 
(T \, = \, 2 \pi / k ) $$
are the periods of potential $\, V_{1} ({\bf r}) \, $.

 It is therefore natural to introduce classes of 
equivalent potentials, assuming
$$V ({\bf r} , \, \alpha , \, {\bf a}) \quad \cong \quad 
V ({\bf r} , \, \alpha , \, {\bf a} + {\bf a}_{pqij}) 
\,\,\, , $$
where
\begin{multline}
\label{apqij}
{\bf a}_{pqij} \quad = \quad  
p \, {\bf e}_{1} \,\, + \,\, q \, {\bf e}_{2} \,\, + \,\,
i \, \pi_{\alpha} \left[ {\bf e}_{1} \right] \,\, + \,\,
j \, \pi_{\alpha} \left[ {\bf e}_{2} \right] \,\,\, ,  \\ 
p, q, i, j \,\, \in \,\, \mathbb{Z} 
\end{multline}

 It is obvious that equivalent potentials have identical 
level lines.

 For generic angles $\, \alpha \, $, the vectors (\ref{apqij}) 
form an everywhere dense set in the space of parameters 
$\, {\bf a} \, $.

 For ``magic'' angles $\, \bar{\alpha}_{n,m} \, $ 
the vectors (\ref{apqij}) form a (rotated) square lattice
with step $\, T / \sqrt{m_{0}^{2} + n_{0}^{2}} \, $.

\vspace{1mm}

 Unlike the quasiperiodic potentials 
$\, V ({\bf r} , \, \alpha , \, {\bf a}) \, $, each of 
the periodic potentials 
$\, V ({\bf r} , \, \bar{\alpha}_{n,m} , \, {\bf a}) \, $ 
has its own interval of open level lines
$$\left[ - \, \epsilon_{n,m} ({\bf a}) \, , \,\, 
\epsilon_{n,m} ({\bf a}) \right] \,\,\, , $$
symmetric with respect to the value $\, \epsilon_{0} = 0 \, $.

\begin{figure}[t]
\begin{center}
\includegraphics[width=\linewidth]{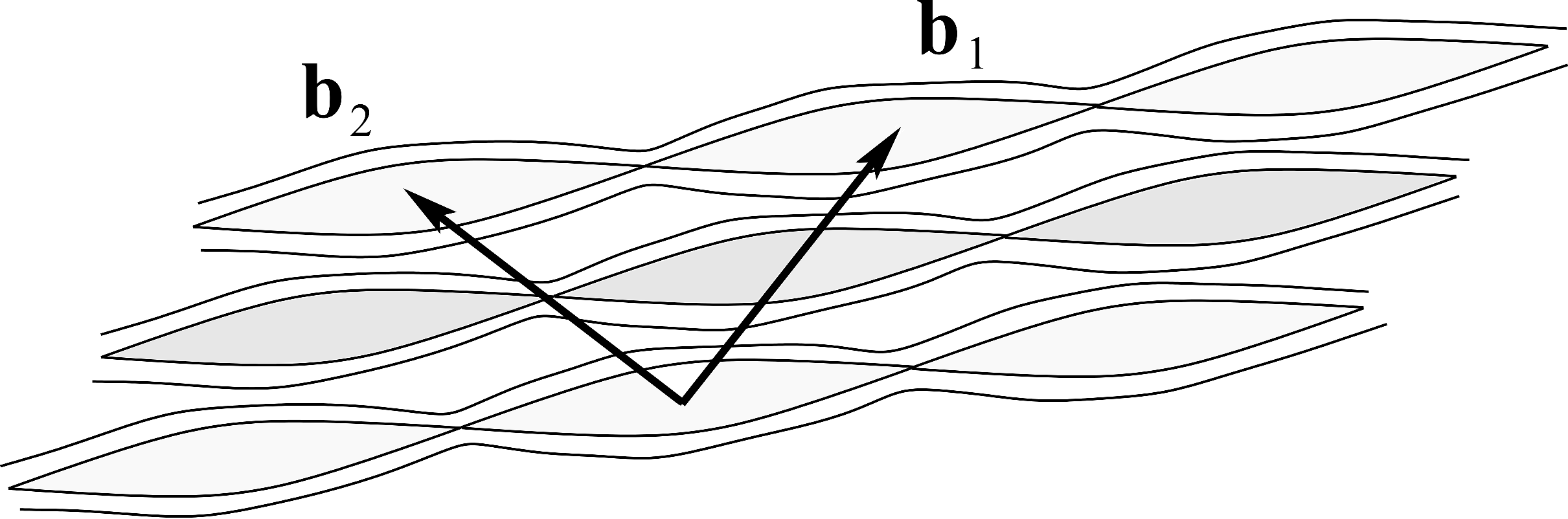}
\end{center}
\caption{Periodic open level lines of the potentials
$\, V ({\bf r} , \, \bar{\alpha}_{n,m} , \, {\bf a}) \, $
(schematically).}
\label{PeriodicLines}
\end{figure}

  Non-singular open level lines 
$\, V ({\bf r} , \, \bar{\alpha}_{n,m} , \, {\bf a}) = \epsilon \, $ 
are periodic and have some common integer direction in the basis 
$\, \left\{ {\bf b}_{1}^{n,m} , \, {\bf b}_{2}^{n,m} \right\} \, $ 
or 
$\, \left\{ {\bf b}_{1}^{\prime \, n,m} , \, 
{\bf b}_{2}^{\prime \, n,m} \right\} \, $ 
(Fig. \ref{PeriodicLines}). These directions, however, 
may differ for different values of $\, {\bf a} \, $.
The situations $\, A_{-} \, $ and $\, A_{+} \, $ for such 
potentials correspond to the values 
$\, \epsilon \, < \, - \, \epsilon_{n,m} ({\bf a}) \, $ 
and $\, \epsilon \, > \, \epsilon_{n,m} ({\bf a}) \, $ 
respectively.

\begin{figure}[t]
\begin{center}
\includegraphics[width=\linewidth]{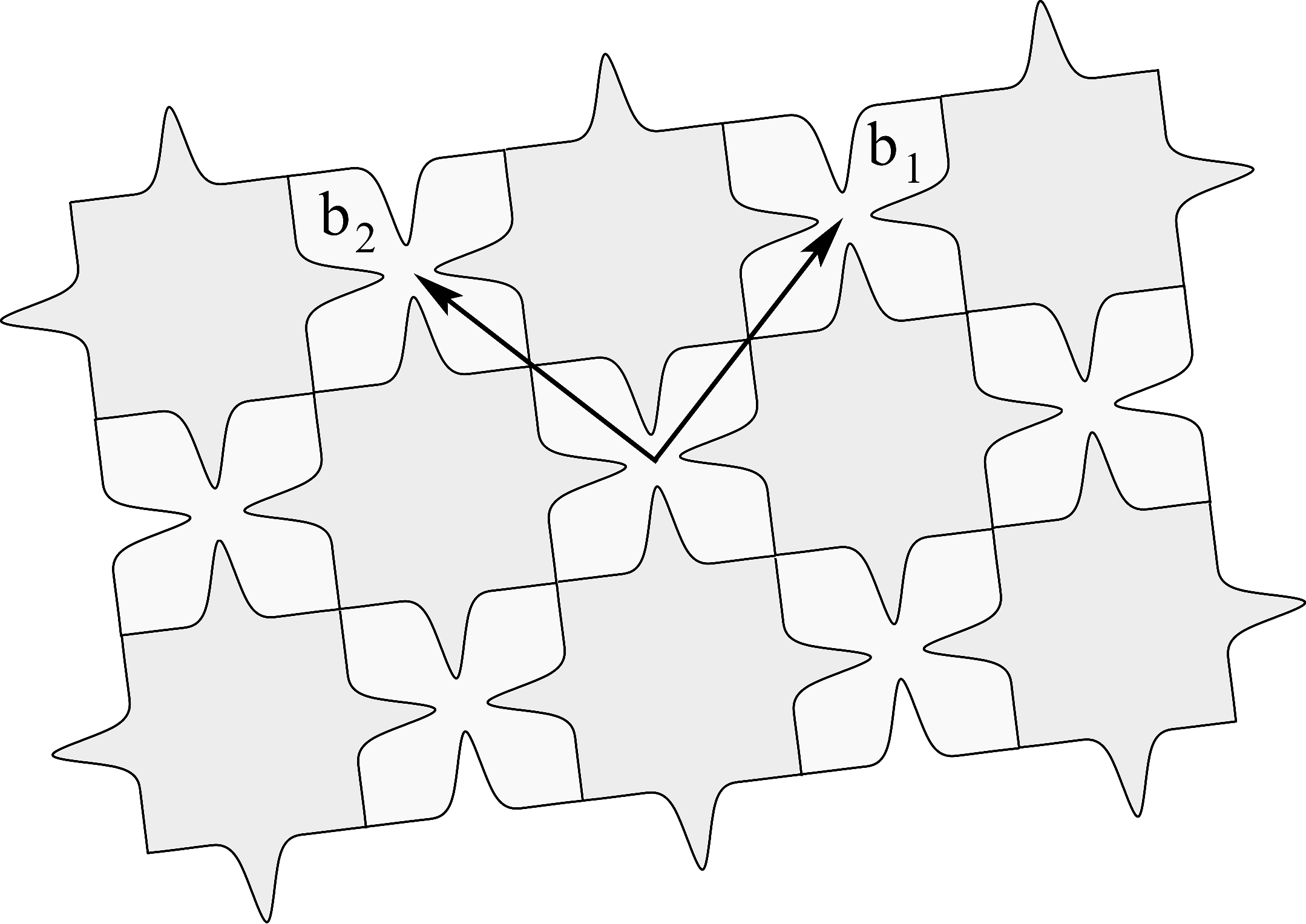}
\end{center}
\caption{``Singular net'' arising at the level 
$\, V ({\bf r} , \, \bar{\alpha}_{n,m} , \, {\bf a}) \, = \, 0 \, $ 
for periodic potentials with exact rotational symmetry (schematically).}
\label{SingularNet}
\end{figure}

 Potentials 
$\, V ({\bf r} , \, \bar{\alpha}_{n,m} , \, {\bf a}) \, $, 
possessing exact rotational symmetry, obviously cannot have 
non-singular periodic open level lines. For such potentials, 
the situations $\, A_{-} \, $ and $\, A_{+} \, $ are separated 
by a (symmetric) periodic ``singular net'' arising at the 
level $\, \epsilon_{0} = 0 \, $ (Fig. \ref{SingularNet}).

 For large values of $\, n \, $ and $\, m \, $, 
the ``singular net'' can have a rather complex geometry. 
In the generic case, however, all such nets are equivalent 
to the net shown in Fig. \ref{SingularNet} from the 
topological point of view, namely, such a net contains 
exactly 2 nonequivalent saddle points of the potential 
$\, V ({\bf r} , \, \bar{\alpha}_{n,m} , \, {\bf a}) \, $, 
and all its ``cells'' have rotational symmetry of the 4th order. 
In addition, the sizes of the ``cells'' of such a net are 
restricted by the formula
\begin{equation}
\label{DEstimation}
T_{n,m} \,\,\, \leq \,\,\,  D  \,\,\, \leq \,\,\, 
\sqrt{2} \,\, T_{n,m} \,\,\, , 
\end{equation}
where 
$$T_{n,m} \,\,\, = \,\,\, T \, \sqrt{m_{0}^{2} + n_{0}^{2}} $$
is the length of the minimal periods of potential
$\, V ({\bf r} , \, \bar{\alpha}_{n,m} , \, {\bf a}) \, $
(we will prove this fact more rigorously in the Appendix).

 All potentials 
$\, V ({\bf r} , \, \bar{\alpha}_{n,m} , \, {\bf 0}) \, $ 
have an infinite number of 4th order symmetry centers located 
at the points
\begin{multline}
\label{SymCenters}
\quad \quad \quad \quad 
{p - q \over 2} \,\, {\bf b}_{1}^{n,m} \,\,\,\,\, + \,\,\,\,\,
{p + q \over 2} \,\, {\bf b}_{2}^{n,m}   \\
\text{or}   \quad \quad
{p - q \over 2} \,\, {\bf b}_{1}^{\prime \, n,m} 
\,\,\,\,\, + \,\,\,\,\,
{p + q \over 2} \,\, {\bf b}_{2}^{\prime \, n,m} 
\quad \quad \quad
\end{multline}
($p, q \, \in \, \mathbb{Z}$). 

 Potentials
\begin{equation}
\label{ijPotentials}
V \left( {\bf r} , \,\, \bar{\alpha}_{n,m} , \,\, 
\pi_{\bar{\alpha}_{n,m}} \left[ 
{i - j \over 2} \, {\bf e}_{1} \,\, + \,\, 
{i + j \over 2} \, {\bf e}_{2} \right] \right) \,\,\, , 
i, j \, \in \, \mathbb{Z} \,\,\, ,
\end{equation}
obviously also have the same symmetry centers.

 Each symmetry center (\ref{SymCenters}) 
has 4 symmetry axes of the potential 
$\, V ({\bf r} , \, \bar{\alpha}_{n,m} , \, {\bf 0}) \, $ 
(as well as potentials (\ref{ijPotentials})) passing 
through it. It is easy 
to see that such symmetry axes form the angles
$${\bar{\alpha}_{n,m} \over 2} \,\, , \,\,\,
{\bar{\alpha}_{n,m} \over 2} \, + \, 45^{\circ} , \,\,\,
{\bar{\alpha}_{n,m} \over 2} \, + \, 90^{\circ}
\quad \text{and} \quad
{\bar{\alpha}_{n,m} \over 2} \, + \, 135^{\circ} $$
with the $\, x $ - axis. 

 As is easy to show, all potentials 
$\, V ({\bf r} , \, \bar{\alpha}_{n,m} , \, {\bf a}) \, $ 
that have exact rotational symmetry have similar (shifted) 
sets of symmetry centers. In general, the set of potentials 
$\, V ({\bf r} , \, \bar{\alpha}_{n,m} , \, {\bf a}) \, $ 
that have exact rotational symmetry is given by the potentials
\begin{equation}
\label{Vprime}
V \left( {\bf r} , \, \bar{\alpha}_{n,m} , \,
{\bf a}^{\prime}_{pqij} \right) \,\,\, , 
\end{equation}
where
\begin{multline}
\label{aprimepqij}
{\bf a}^{\prime}_{pqij} \quad = \quad 
{p - q \over 2} \,\, {\bf e}_{1} \,\,\, + \,\,\, 
{p + q \over 2} \,\, {\bf e}_{2} \,\,\, +  \\
+ \,\,\,
{i - j \over 2} \,\, \pi_{\bar{\alpha}_{n,m}} \left[ 
{\bf e}_{1} \right] \,\,\, + \,\,\,
{i + j \over 2} \,\, \pi_{\bar{\alpha}_{n,m}} \left[ 
{\bf e}_{2} \right] \,\,\, , 
\end{multline}
$p, q, i, j \, \in \, \mathbb{Z} \, $. 

 The vectors (\ref{aprimepqij}) form a square lattice 
in the space of parameters $\, {\bf a} \, $ with the 
step $\, T / 2 \sqrt{m_{0}^{2} + n_{0}^{2}} \, $. 
The lattice (\ref{aprimepqij}) contains the lattice 
(\ref{apqij}) and is ``denser''. In particular, 
for any potential 
$\, V ({\bf r} , \, \bar{\alpha}_{n,m} , \, {\bf a}) \, $ 
there always exists a potential (\ref{Vprime}) such that
$$\left| {\bf a} \, - \, {\bf a}^{\prime}_{pqij} \right|
\,\,\, \leq \,\,\, 
{T \over 2 \sqrt{2 (m_{0}^{2} + n_{0}^{2})}} $$

 The above relation also allows us to prove the relations
$$\epsilon_{n,m} ({\bf a}) \,\,\, \leq \,\,\, 
{\pi \, V_{0} \over \sqrt{m_{0}^{2} + n_{0}^{2}}} $$
for the values $\, \epsilon_{n,m} ({\bf a}) \, $.

 Indeed, using the obvious relation
\begin{equation}
\label{nablaV}
\big| \nabla_{{\bf a}} 
V \left ({\bf r} , \, \alpha , \, {\bf a} \right) \big|
\,\,\, \leq \,\,\, \sqrt{2} \, k \, V_{0} \,\,\, , 
\end{equation}
we get
\begin{multline*}
\big|  V ({\bf r} , \, \bar{\alpha}_{n,m} , \, {\bf a})
\, - \,  V \left( {\bf r} , \, \bar{\alpha}_{n,m} , \,
{\bf a}^{\prime}_{pqij} \right) \big| \,\,\, \leq  \\
\leq \,\,\, 
{ \sqrt{2} \, k \, V_{0} \, T  \over 
2 \sqrt{2 (m_{0}^{2} + n_{0}^{2})}} \,\,\, = \,\,\, 
{\pi \, V_{0} \over \sqrt{m_{0}^{2} + n_{0}^{2}}} 
\end{multline*}

 Thus, for any
$$\epsilon \,\,\, > \,\,\, 
{\pi \, V_{0} \over \sqrt{m_{0}^{2} + n_{0}^{2}}} $$
the set
$$ V ({\bf r} , \, \bar{\alpha}_{n,m} , \, {\bf a})
\,\,\, > \,\,\, \epsilon $$
is contained within the set
$$V \left( {\bf r} , \, \bar{\alpha}_{n,m} , \,
{\bf a}^{\prime}_{pqij} \right) \,\,\, > \,\,\, 0 $$

 The latter set, in turn, has only bounded components 
(lying inside the cells of the ``singular net''), 
and thus for the potential 
$\, V ({\bf r} , \, \bar{\alpha}_{n,m} , \, {\bf a}) \, $ 
we have in this case a situation of the type $\, A_{+} \, $.

 The reasoning is completely similar for the values
$$\epsilon \,\,\, < \,\,\, - \,  
{\pi \, V_{0} \over \sqrt{m_{0}^{2} + n_{0}^{2}}} $$

 Note here also that for generic angles $\, \alpha \, $ 
(not ``magic'') the potentials 
$\, V ({\bf r} , \, \alpha , \, {\bf a}) \, $, which have 
exact rotational symmetry of order 4 (with a single center 
of symmetry), represent an everywhere dense set among all 
potentials $\, V ({\bf r} , \, \alpha , \, {\bf a}) \, $.

\section{Potential 
$\, V ({\bf r} , \, 45^{\circ} , \, {\bf 0}) \, $ 
and related potentials}
\setcounter{equation}{0}

 Here we will consider in more detail the potential 
$\, V ({\bf r} , \, 45^{\circ} , \, {\bf 0}) \, $, 
which has exact rotational symmetry of the 8th order, 
as well as the potentials 
$$ V ({\bf r} , \, 45^{\circ} , \, {\bf a}) \,\,\, ,
\quad  {\bf a} \in \mathbb{R}^{2} $$
associated with it.

 We will need an approximation of the potentials 
$\, V ({\bf r} , \, 45^{\circ} , \, {\bf a}) \, $ 
by periodic potentials, which, as we have already said, 
is associated with approximations of the quantity
$$\tan \, 22.5^{\circ} \,\,\, = \,\,\, 
\sqrt{2} \,\, - \,\, 1 $$
by rational fractions.

 The value $\, \sqrt{2} - 1 \, $ has the following 
expansion into a continued fraction
$$\sqrt{2} \, - \, 1  \quad = \quad 
\cfrac{1}{2 + \cfrac{1}{2 +  \cfrac{1}{2 + \dots}}} $$

 Successive reductions of the continued fraction give 
approximations $\, m^{(s)} / n^{(s)} \, $ for 
$\, \sqrt{2} - 1 \, $, and it can be shown that
$$m^{(s+1)} \,\,\, = \,\,\, n^{(s)} \,\,\, , \quad \quad
n^{(s+1)} \,\,\, = \,\,\, m^{(s)} \,\, + \,\, 2 \, n^{(s)}
\,\,\, , $$
such that 
$${m^{(s)} \over n^{(s)}} \quad \rightarrow \quad 
{m^{(s+1)} \over n^{(s+1)}} \quad  =  \quad 
{n^{(s)} \over m^{(s)} + 2 n^{(s)}} $$

 Assuming $\, m^{(1)} = 1 \, $, $\, n^{(1)} = 2 \, $, 
it is easy to see that the numbers $\, m^{(s)} \, $ and 
$\, n^{(s)} \, $ are relatively prime and have different 
parities for any $\, s \, $.

 We can also write
$$\left( 
\begin{array}{c}
m^{(s+1)}  \\
n^{(s+1)}
\end{array}  \right)     \quad  =  \quad 
\left( 
\begin{array}{cc}
0  &  1   \\
1  &  2  
\end{array}  \right) \, 
\left( 
\begin{array}{c}
m^{(s)}  \\
n^{(s)}
\end{array}  \right)   $$

 The eigenvalues of the above system are
$$\lambda_{1} \,\,\, = \,\,\, 1 \, - \, \sqrt{2} 
\,\,\, , \quad \quad
\lambda_{2} \,\,\, = \,\,\, 1 \, + \, \sqrt{2}
\,\,\, , $$
and the eigenvectors can be chosen as
$$\bm{\xi}_{1} \,\,\, = \,\,\,
\left( 
\begin{array}{c}
1  \\
1 - \sqrt{2} 
\end{array}  \right) \,\,\, , \quad \quad
\bm{\xi}_{2} \,\,\, = \,\,\,
\left( 
\begin{array}{c}
1  \\
\sqrt{2} + 1
\end{array}  \right)   $$

 We have then
\begin{multline*}
\left( 
\begin{array}{c}
m^{(1)}  \\
n^{(1)}
\end{array}  \right)     \quad  =  \quad 
\left( 
\begin{array}{c}
1  \\
2
\end{array}  \right)      
\quad  =    \\
= \quad 
{\sqrt{2} - 1 \over 2 \sqrt{2}}
\left( 
\begin{array}{c}
1  \\
1 - \sqrt{2} 
\end{array}  \right) \,\,\, + \,\,\, 
{\sqrt{2} + 1 \over 2 \sqrt{2}}
\left( 
\begin{array}{c}
1  \\
\sqrt{2} + 1
\end{array}  \right)   
\end{multline*}
and 
$$m^{(s)} \,\,\, = \,\,\, {1 \over 2 \sqrt{2}} \,
\Big( (-1)^{s-1} \left( \sqrt{2} - 1 \right)^{s} 
\,\, + \,\, \left( \sqrt{2} + 1 \right)^{s} \Big) $$
$$n^{(s)} \,\,\, = \,\,\, {1 \over 2 \sqrt{2}} \,
\Big( (-1)^{s} \left( \sqrt{2} - 1 \right)^{s+1} 
\,\, + \,\, \left( \sqrt{2} + 1 \right)^{s+1} \Big) $$
$${m^{(s)} \over n^{(s)}} \quad = \quad 
{\sqrt{2} \,  - \,  1 \,\,\, + \,\,\, (-1)^{s-1} 
\left( \sqrt{2} - 1 \right)^{2s+1}  \over
1 \,\,\, + \,\,\, (-1)^{s} 
\left( \sqrt{2} - 1 \right)^{2s+2}} $$

 It can be seen that (according to the general theory) 
fractions $\, m^{(s)} / n^{(s)} \, $ with even and odd 
$\, s \, $ approach  \linebreak
$\, \sqrt{2} - 1 \, $ ``from different sides''.

 For large values of $\, s \, $ we can put with 
good accuracy
$$m^{(s)} \,\,\, \simeq \,\,\,
{\left( \sqrt{2} + 1 \right)^{s} \over 2 \sqrt{2}}
\,\,\, , \quad \quad
n^{(s)} \,\,\, \simeq \,\,\,
{\left( \sqrt{2} + 1 \right)^{s+1} \over 2 \sqrt{2}} $$
\begin{multline*}
{m^{(s)} \over n^{(s)}} \,\,\, \simeq \,\,\,
\left( \sqrt{2} - 1 \right)  \, \times  \\
\times  \left( 1 \,\, + \,\,
(-1)^{s-1} \left( \left( \sqrt{2} - 1 \right)^{2s}
\, + \, \left( \sqrt{2} - 1 \right)^{2s+2} \right)
\right) \,\, =  \\
= \,\, \left( \sqrt{2} - 1 \right) 
\left( 1 \,\,\, + \,\,\,
(-1)^{s-1} \,\, 2 \sqrt{2} \left(\sqrt{2} - 1 \right)^{2s+1}
\right)
\end{multline*}

 For even $\, s\, $ we have also the strict inequality
\begin{multline*}
\sqrt{2} \,  - \,  1  \quad > \quad {m^{(s)} \over n^{(s)}}
\quad >  \\
> \quad \left( \sqrt{2} - 1 \right) 
\left( 1 \,\,\, - \,\,\,
2 \sqrt{2} \left(\sqrt{2} - 1 \right)^{2s+1} \right)
\end{multline*}

 As we have already noted, the numbers $\, m^{(s)} \, $ 
and $\, n^{(s)} \, $ have different parities for any 
$\, s \, $, therefore the length of the minimal period 
$\, T_{(s)} \, $ of the potentials
$$V \left( {\bf r} , \, \bar{\alpha}_{n^{(s)},m^{(s)}} , \,
{\bf a} \right)  \quad  \equiv  \quad 
V \left( {\bf r} , \, \bar{\alpha}_{(s)} , \, {\bf a} \right) $$
is equal to
\begin{multline*}
T_{(s)}  \quad  =  \quad  T \, 
\sqrt{\left( m^{(s)} \right)^{2} + \left( n^{(s)} \right)^{2}} 
\quad  =  \\
=  \,
T \,\, {\left( \sqrt{2} + 1 \right)^{s} \over 2 \sqrt{2}} \, 
\sqrt{4 + 2 \sqrt{2} \, + \,
\left(\sqrt{2} - 1 \right)^{4s} 
\left( 4 - 2 \sqrt{2} \right)}   =   \\
=  \quad  
T \,\, {\left( \sqrt{2} + 1 \right)^{s + 1/2} \over 
\sqrt{2 \sqrt{2}}} \,\, 
\sqrt{1 \, + \, \left(\sqrt{2} - 1 \right)^{4s+2} } 
\end{multline*}

 Thus, for all $\, s \, $ we have
\begin{multline*}
\quad T \,\, {\left( \sqrt{2} + 1 \right)^{s + 1/2} \over 
\sqrt{2 \sqrt{2}}} \quad  <  \quad  T_{(s)}  \quad  <   \\
<  \quad T \,\, {\left( \sqrt{2} + 1 \right)^{s + 1/2} \over 
\sqrt{2 \sqrt{2}}} \,\,\, + \,\,\, 
T \,\, {\left( \sqrt{2} - 1 \right)^{3s + 3/2} \over 
2 \sqrt{2 \sqrt{2}}} 
\end{multline*}

 Using the relation (\ref{tanSootn}), we can put with 
good accuracy for $\, s \gg 1 \, $
\begin{multline*}
\bar{\alpha}_{(s)} \, - \, 45^{\circ} \,\,\, \simeq \,\,\, 
\left( {m^{(s)} \over n^{(s)}} \,\, - \,\, \left(
\sqrt{2} - 1 \right) \right) {\sqrt{2} + 1 \over \sqrt{2}} 
\,\,\, \simeq   \\
\simeq \,\,\, (-1)^{s-1} \,\,\, 2 \sqrt{2}  
\left(\sqrt{2} - 1 \right)^{2s+2} {\sqrt{2} + 1 \over \sqrt{2}}
\,\,\, =   \\
= \,\,\, (-1)^{s-1} \,\, 2  \left(\sqrt{2} - 1 \right)^{2s+1}
\end{multline*}
 
 For even $\, s\, $ we have also the strict inequality
$$- \, 2  \left(\sqrt{2} - 1 \right)^{2s+1} \quad < \quad 
\bar{\alpha}_{(s)} \, - \, 45^{\circ} \quad < \quad 0 $$
 
 Similarly, with very good accuracy for large $\, s \, $, 
we can write
$$\left| \bar{\alpha}_{(s)} \, - \, 45^{\circ} \right| 
\, \cdot \, \sqrt{2} \, T_{(s)} \quad \simeq  \quad 
\sqrt{2 \sqrt{2}} \,\,\, T \,\, 
\left(\sqrt{2} - 1 \right)^{s + 1/2} $$

 The above relations allow us to estimate the growth rate 
of the sizes of closed level lines of
$\, V ({\bf r} , \, 45^{\circ} , \, {\bf a}) \, $ as 
$\, \epsilon \, $ approaches zero. As we have already noted, 
the potentials equivalent to the potential 
$\, V ({\bf r} , \, 45^{\circ} , \, {\bf 0}) \, $ 
correspond to an everywhere dense set among all 
$\, {\bf a} \in \mathbb{R}^{2} \, $, so it is natural 
not to single out here the potential 
$\, V ({\bf r} , \, 45^{\circ} , \, {\bf 0}) \, $ 
and to consider the entire family 
$\, V ({\bf r} , \, 45^{\circ} , \, {\bf a}) \, $ at once.

 Thus, the above relations imply the existence 
of a sequence of values
\begin{multline*}
\epsilon_{(s)} \quad = \quad  \sqrt{2} k V_{0} \,\,
\cdot \left| \bar{\alpha}_{(s)} \, - \, 45^{\circ} \right| 
\, \cdot \, \sqrt{2} \, T_{(s)} \quad +  \\
+ \quad { \sqrt{2} k V_{0} \cdot T \over 
2 \sqrt{2 \left( \left( m^{(s)} \right)^{2} + 
\left( n^{(s)} \right)^{2} \right)}}  \quad \simeq   \\
\simeq \quad  2^{9/4} \pi V_{0} 
\left(\sqrt{2} - 1 \right)^{s + 1/2} \,\,\, + \,\,\,
{\pi V_{0} \sqrt{2 \sqrt{2}} \over 
\left(\sqrt{2} + 1 \right)^{s + 1/2}}  \quad  =  \\
=  \quad   \pi V_{0} \, 
\left(  2^{9/4} \, + \,  2^{3/4} \right)
\left(\sqrt{2} - 1 \right)^{s + 1/2}
\end{multline*}
such that for any $\, | \epsilon | > \epsilon_{(s)} \, $, 
the size of the level lines
\begin{equation}
\label{V45epsilon}
V ({\bf r} , \, 45^{\circ} , \, {\bf a}) 
\,\,\, = \,\,\, \epsilon 
\end{equation}
does not exceed
\begin{multline*}
\sqrt{2} \, T_{(s)}  \,\,\, \simeq \,\,\, 
T \,\, 2^{-1/4} \, \left( \sqrt{2} + 1 \right)^{s + 1/2}  
\,\,\, \simeq  \\
\simeq  \,\,\, \pi \, V_{0} \, T \,\, 
\left( 4 + \sqrt{2} \right) \,\, \epsilon_{(s)}^{-1} 
\end{multline*}

 Indeed, let, for example, 
$\, \epsilon \, < \, - \, \epsilon_{(s)} \, $ and the 
set (\ref{V45epsilon}) contain a connected component 
passing through some point $\, (x_{0} , y_{0}) \, $. 
Consider the potential 
$\, \widetilde{V}^{(s)}_{(x_{0} , y_{0})} ({\bf r}, \, {\bf a}) \, $, 
formed by the superposition of the potential 
$\, V_{1} ({\bf r}) \, $ and the potential 
$\, V_{2} ({\bf r}) \, $, rotated by the angle
$$\delta \alpha_{(s)} \,\,\, = \,\,\, 
\bar{\alpha}_{(s)} \,\, - \,\, 45^{\circ} $$
relative to the point $\, (x_{0} , y_{0}) \, $.
Obviously
\begin{equation}
\label{widetildeV}
\widetilde{V}^{(s)}_{(x_{0} , y_{0})}
({\bf r}, \, {\bf a}) \,\,\, = \,\,\, 
V \left( {\bf r} , \, \bar{\alpha}_{(s)} , \, 
{\bf a}^{\prime} \right) 
\end{equation}
for some $\, {\bf a}^{\prime} \, $.

 In the circle of radius $\, \sqrt{2} \, T_{(s)} \, $ 
with center $\, (x_{0} , y_{0}) \, $ we obviously have 
for the initial potential
$\, V ({\bf r} , \, 45^{\circ} , \, {\bf a}) \, $:
\begin{multline*}
\left| V ({\bf r} , \, 45^{\circ} , \, {\bf a})
\, - \, \widetilde{V}^{(s)}_{(x_{0} , y_{0})}
({\bf r}, \, {\bf a}) \right| \,\,\, \leq  \\
\leq \,\,\, \sqrt{2} k V_{0} \,\,
\cdot \left| \bar{\alpha}_{(s)} \, - \, 45^{\circ} 
\right| \, \cdot \, \sqrt{2} \, T_{(s)}
\end{multline*}

 According to what we have said above, for the 
potential (\ref{widetildeV}) there exists a potential
$$\widehat{V}^{(s)}_{(x_{0} , y_{0})}
({\bf r}, \, {\bf a}) \,\,\, = \,\,\, 
V \left( {\bf r} , \, \bar{\alpha}_{(s)} , \, 
{\bf a}^{\prime\prime} \right) \,\,\, , $$
possessing exact rotational symmetry (of the 4th order) 
and such that
$$\left| {\bf a}^{\prime} \, - \, {\bf a}^{\prime\prime} 
\right| \,\,\, \leq \,\,\, 
{T \over 2 \sqrt{2 \left( \left( m^{(s)} \right)^{2} + 
\left( n^{(s)} \right)^{2} \right)}} $$
  
 Using the relation (\ref{nablaV}), we can then also write
\begin{multline*}
\left| \widetilde{V}^{(s)}_{(x_{0} , y_{0})}
({\bf r}, \, {\bf a}) \, - \, 
\widehat{V}^{(s)}_{(x_{0} , y_{0})}
({\bf r}, \, {\bf a}) \right|  \quad \leq  \\
\leq  \quad  \sqrt{2} k V_{0} \,\,\, 
{T \over 2 \sqrt{2 \left( \left( m^{(s)} \right)^{2} + 
\left( n^{(s)} \right)^{2} \right)}}
\end{multline*}
and, so, in the circle of radius $\, \sqrt{2} \, T_{(s)} \, $ 
with center $\, (x_{0} , y_{0}) \, $:
\begin{multline*}
\left| V ({\bf r} , \, 45^{\circ} , \, {\bf a})
\, - \, \widehat{V}^{(s)}_{(x_{0} , y_{0})}
({\bf r}, \, {\bf a}) \right|  \quad \leq  \\
\leq  \quad  \sqrt{2} k V_{0} \,\,
\cdot \left| \bar{\alpha}_{(s)} \, - \, 45^{\circ} \right| 
\, \cdot \, \sqrt{2} \, T_{(s)} \quad +  \\
+ \quad { \sqrt{2} k V_{0} \cdot T \over 
2 \sqrt{2 \left( \left( m^{(s)} \right)^{2} + 
\left( n^{(s)} \right)^{2} \right)}} 
\end{multline*}

 It can be seen, therefore, that for 
$\epsilon < - \epsilon_{(s)} $ a connected component 
(\ref{V45epsilon}) passing through $\, (x_{0} , y_{0}) \, $, 
in a circle of radius $\, \sqrt{2} \, T_{(s)} \, $ centered in 
$\, (x_{0} , y_{0}) \, $, also lies in the region
$$\widehat{V}^{(s)}_{(x_{0} , y_{0})}
({\bf r}, \, {\bf a})  \quad  <  \quad  0 $$

 Thus, our component (\ref{V45epsilon}) is entirely 
contained in one of the cells of the ``singular net'' 
of the potential 
$\, \widehat{V}^{(s)}_{(x_{0} , y_{0})}({\bf r}, \, {\bf a}) \, $, 
which imposes a constraint 
$\, D \, \leq \, \sqrt{2} \, T_{(s)} \, $ on its maximal size. 
It is easy to see that under the same assumptions we also have 
the same constraint on the sizes of the regions
$$V ({\bf r} , \, 45^{\circ} , \, {\bf a}) 
\,\,\, < \,\,\, \epsilon $$
(the reasoning for $\, \epsilon > \epsilon_{(s)} \, $ 
is similar to that given above).

 Breaking the full energy range into intervals
$$\left[ - \, \epsilon_{(s)} , \, - \, \epsilon_{(s+1)}
\right] \,\,\, ,  \quad \quad
\left[ \epsilon_{(s+1)} , \,  \epsilon_{(s)}
\right] \,\,\, , $$
and taking into account that in each of the intervals
$$| \epsilon | \,\,\, \leq \,\,\, \epsilon_{(s)}
\,\,\, \simeq \,\,\, \left(\sqrt{2} + 1 \right) \,
\epsilon_{(s+1)} \,\,\, , $$
we can also write the general estimate
\begin{equation}
\label{GenEpsilonEst}
D (\epsilon)  \quad  \leq  \quad  
\pi \, V_{0} \, T \,\, \left( 4 + \sqrt{2} \right) \,\, 
\left(\sqrt{2} + 1 \right) \,\, \epsilon^{-1} 
\end{equation}
for the sizes of level lines (\ref{V45epsilon}) near the 
zero value of $\, \epsilon \, $ ($\epsilon \ll V_{0}$).

\vspace{2mm}

 In conclusion, we note here that the potentials 
$\, V ({\bf r} , \, 45^{\circ} , \, {\bf a}) \, $ are 
of particular interest in many experimental studies of 
two-dimensional systems. This circumstance is due, 
in particular, to the presence of rotational symmetry 
of the 8th order in an everywhere dense subset of the 
potentials of this family.

 At the same time, more general potentials 
$\, V ({\bf r} , \, \alpha , \, {\bf a}) \, $ also have 
many interesting properties and can be distinguished 
among the potentials defined by superpositions of periodic 
potentials on the plane. In particular, such potentials 
also cannot have topologically regular open level lines, 
which may be of some importance from the experimental 
point of view. As we have already noted, open level lines 
of potentials $\, V ({\bf r} , \, \alpha , \, {\bf a}) \, $ 
(for non-``magic'' angles $\, \alpha$) can arise only for 
$\, \epsilon = 0 \, $, while the sizes of connected level 
lines are bounded by a certain constant $\, D (\epsilon) \, $ 
for any other $\, \epsilon \, $.

 Many properties of potentials 
$\, V ({\bf r} , \, \alpha , \, {\bf a}) \, $ bring them close 
to random potentials on the plane, however, like potentials 
$\, V ({\bf r} , \, 45^{\circ} , \, {\bf a}) \, $, they have 
their own distinctive features. In particular, one of such 
feature here is also a slower growth of the sizes of closed level 
lines near $\, \epsilon = 0 \, $, as for potentials 
$\, V ({\bf r} , \, 45^{\circ} , \, {\bf a}) \, $.

 The analysis of the behavior of level lines of the potentials 
$\, V ({\bf r} , \, \alpha , \, {\bf a}) \, $ near 
$\, \epsilon = 0 \, $ largely repeats similar reasoning 
for the potentials 
$\, V ({\bf r} , \, 45^{\circ} , \, {\bf a}) \, $. 
Here we note only one feature that can arise in the most 
general case. Namely, the approximation of 
$\, \tan 45^{\circ} \, $ by rational fractions 
has a fairly ``regular'' form, which, in particular, 
allows us to derive the general estimate (\ref{GenEpsilonEst}). 
For most angles $\, \alpha \, $ (the set of full measure in the 
angle space), such approximations have similar properties, 
which allows us to obtain estimates for them close to 
(\ref{GenEpsilonEst}). Some $\, \tan \alpha \, $, however, 
are approximated by the numbers $\, m^{(s)} / n^{(s)} \, $ 
``too well'', while the numbers $\, m^{(s)} \, $ and 
$\, n^{(s)} \, $ grow ``too fast''. As a consequence, 
a common estimate (\ref{GenEpsilonEst}) for them may be 
absent, and the value $\, D (\epsilon) \, $ may 
have a pronounced ``cascade'' growth. Note that the 
set of corresponding $\, \alpha \, $ has zero measure
in the angle space.

\section{Appendix}
\setcounter{equation}{0}

 Here we prove the relation (\ref{DEstimation}) for the 
diameter of the cells of a ``singular net'' of potentials 
$\, V ({\bf r} , \, \bar{\alpha}_{n,m} , \, {\bf a}) \, $, 
possessing exact rotational symmetry of the 4th order.

\vspace{1mm}

 As we have already said, we will assume here that 
singular nets 
$\, V ({\bf r} , \, \bar{\alpha}_{n,m} , \, {\bf a}) \, $ 
are generic nets from the topological point of view. 
We will consider each cell of such a net as a simply 
connected region $\, \Omega \, $ (ignoring possible 
closed level lines inside it), possessing rotational 
symmetry of the 4th order, as well as reflection 
symmetry of the potential 
$\, V ({\bf r} , \, \bar{\alpha}_{n,m} , \, {\bf a}) \, $.

\vspace{1mm}

 The boundary of $\, \Omega \, $ contains 4 saddle points 
of the potential 
$\, V ({\bf r} , \, \bar{\alpha}_{n,m} , \, {\bf a}) \, $, 
representing two pairs of equivalent (differing by a shift 
by a period of 
$\, V ({\bf r} , \, \bar{\alpha}_{n,m} , \, {\bf a})$) 
saddle singular points (Fig. \ref{SingularNet}). 
It is easy to see that the distance between diametrically 
opposite (equivalent) saddle points on the boundary of 
$\, \Omega \, $ is equal to $\, T_{n,m} \, $, so that in 
any case we have the relation $\, D \, \geq \, T_{n,m} \, $.

\vspace{1mm}

\begin{figure}[t]
\begin{center}
\includegraphics[width=\linewidth]{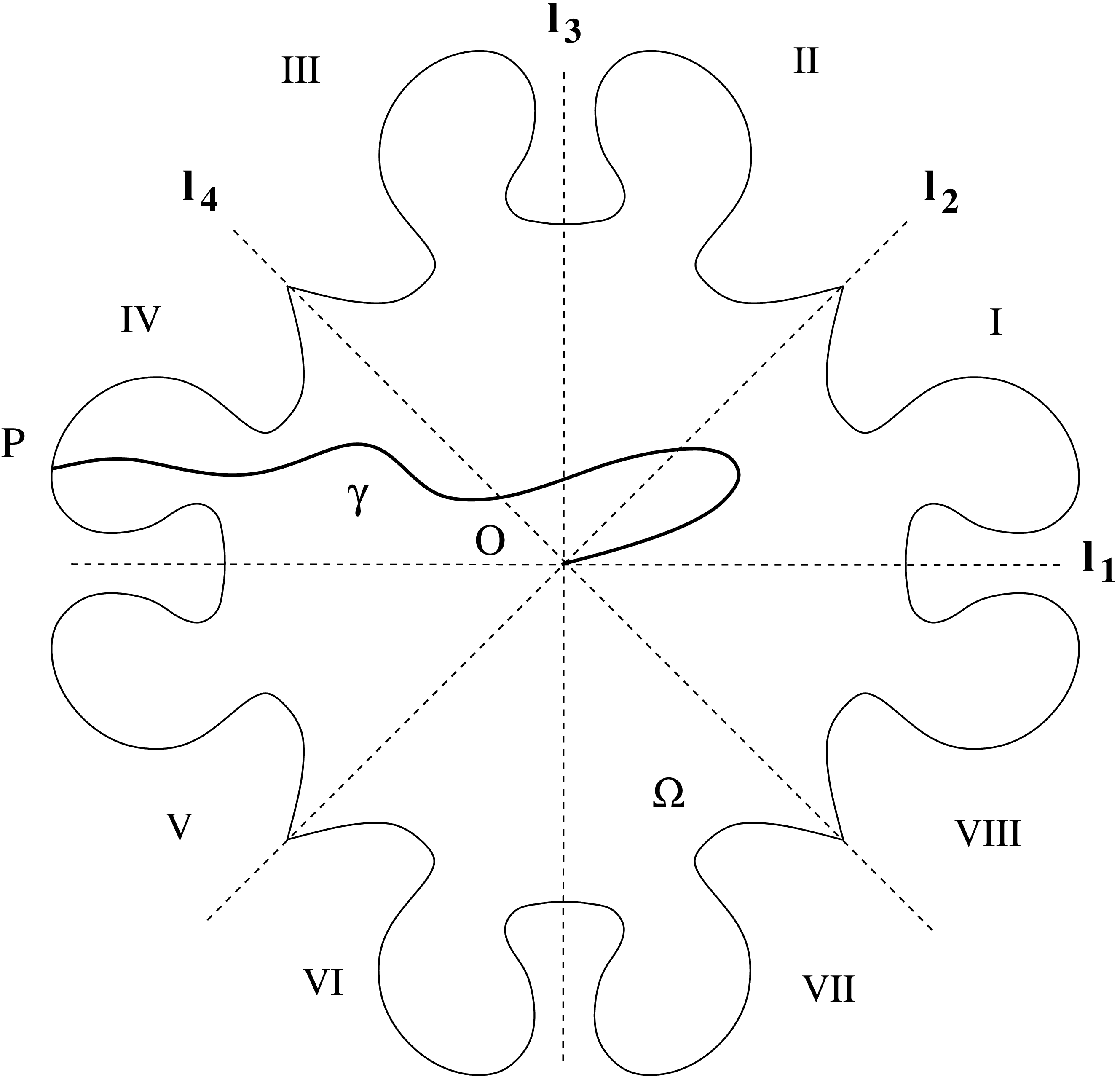}
\end{center}
\caption{A cell of a ``singular net'' of potential 
$\, V ({\bf r} , \, \bar{\alpha}_{n,m} , \, {\bf a}) \, $, 
possessing exact rotational symmetry (schematically).}
\label{SingDomain}
\end{figure}

 The center of $\, \Omega \, $ has 4 symmetry axes
$\, {\bf l}_{1} \, $, $\, {\bf l}_{2} \, $, 
$\, {\bf l}_{3} \, $, $\, {\bf l}_{4} \, $
passing through it, which divide $\, \mathbb{R}^{2} \, $ 
into 8 sectors (octants) I - VIII (Fig. \ref{SingDomain}).
Let $\, \gamma \subset \Omega \, $ be some curve connecting 
the center of the domain $\, \Omega \, $ with the point 
$\, P \, $ that is most distant from it and lies on the 
boundary of $\, \Omega \, $ (Fig. \ref{SingDomain}). 
Obviously, $\, D \, = \, 2 \left| O P \right| \, $.

\vspace{1mm}

\begin{figure}[t]
\begin{center}
\includegraphics[width=\linewidth]{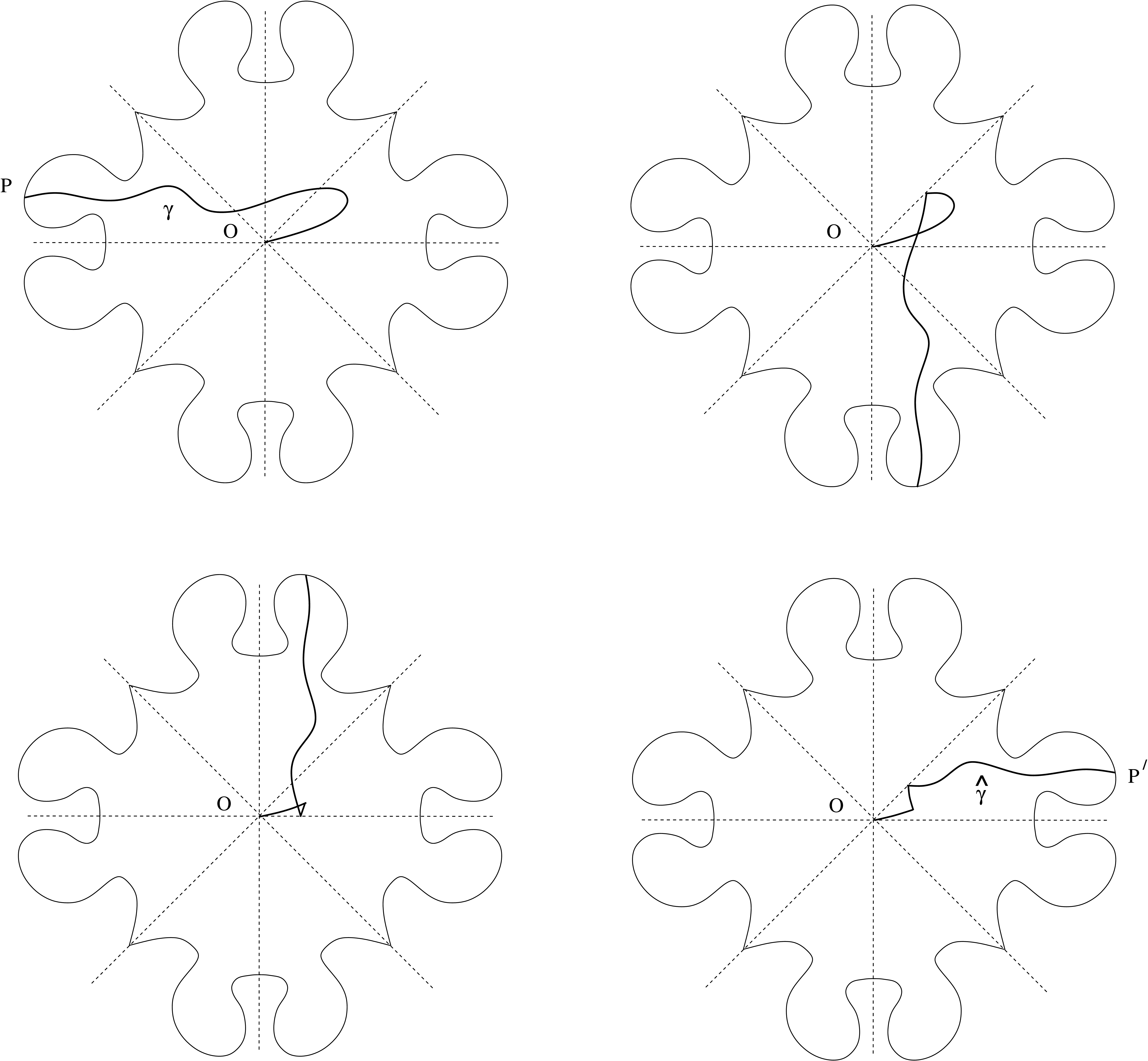}
\end{center}
\caption{Curve $\, \widehat{\gamma} \, $ in the region 
$\, \Omega \, $. } 
\label{gammahat}
\end{figure}

 Without loss of generality, let the initial velocity 
vector on the curve $\, \gamma \, $ lie in the octant I. 
Using reflections with respect to the symmetry axes, 
as well as reconstructions of $\, \gamma \, $, we can 
construct a curve $\, \widehat{\gamma} \, $ that lies 
entirely in the octant I and connects the point $\, O \, $ 
with a point $\, P^{\prime} \in \partial \Omega \, $ 
such that 
$\, \left| O P \right| = \left| O P^{\prime} \right| \, $ 
(Fig. \ref{gammahat}). It is also easy to see that 
by a small perturbation the curve $\, \widehat{\gamma} \, $ 
can be made a smooth curve, all of whose interior points 
lie inside the octant I.

\vspace{1mm}

 Either $\, {\bf l}_{1} \, $ or $\, {\bf l}_{2} \, $ must 
contain a symmetry center $\, O^{\prime} \, $ obtained 
from the point $\, O \, $ by a shift by the minimal period 
$\, {\bf T} \, $ of the potential 
$\, V ({\bf r} , \, \bar{\alpha}_{n,m} , \, {\bf a}) \, $. 
Let (without loss of generality) this be the symmetry axis 
$\, {\bf l}_{1} \, $.

\vspace{1mm}

\begin{figure*}
\begin{center}
\includegraphics[width=175mm]{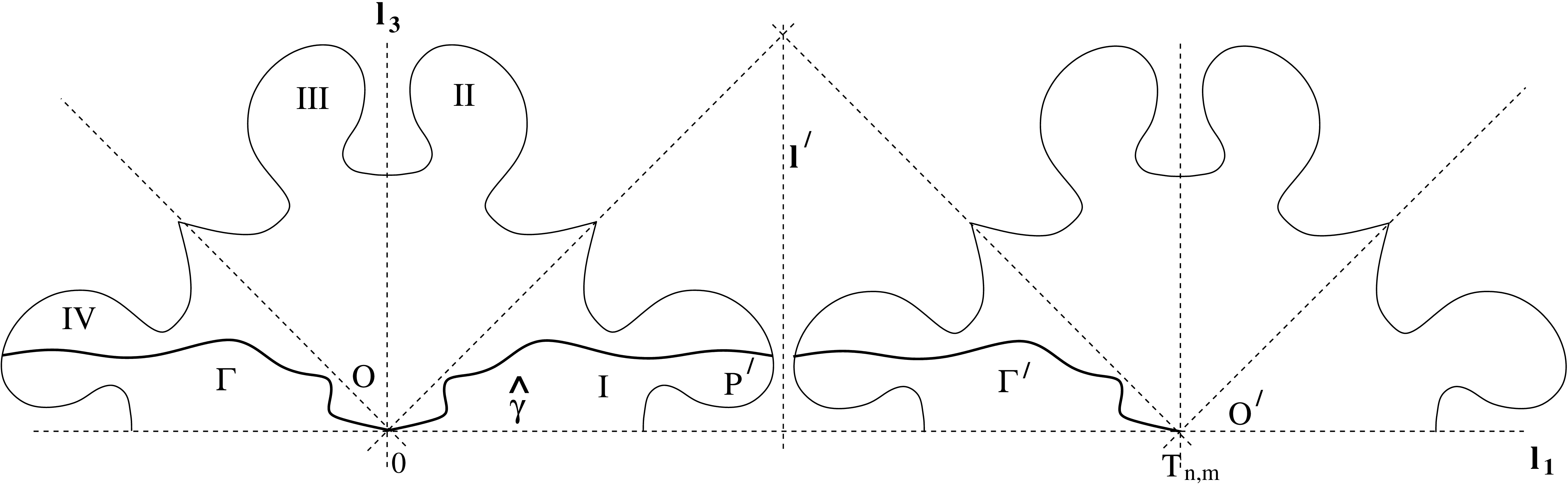}
\end{center}
\caption{Curves $\, \widehat{\gamma} \, $, $\, \Gamma \, $ 
and $\, \Gamma^{\prime} \, $ in the plane 
$\, \mathbb{R}^{2} \, $.}
\label{gammaGamma}
\end{figure*}

 Consider the curve $\, \Gamma \subset \Omega \, $ obtained 
from $\, \widehat{\gamma} \, $ by reflection about the axis 
$\, {\bf l}_{3} \, $ and lying in octant IV. The shift of 
$\, \Gamma \, $ by the period $\, {\bf T} \, $ lies in 
another cell of the ``singular net'' of 
$\, V ({\bf r} , \, \bar{\alpha}_{n,m} , \, {\bf a}) \, $ 
and, thus, should not intersect $\, \widehat{\gamma} \, $ 
at interior points. However, since
$$\Gamma^{\prime} \quad = \quad \Gamma \,\,\, + \,\,\, 
{\bf T} $$
is the reflection of $\, \widehat{\gamma} \, $ about the 
symmetry axis $\, {\bf l}^{\prime} \, $ 
(Fig. \ref{gammaGamma}), this is possible only for
$$\left| O P^{\prime} \right| \,\,\, \leq \,\,\, 
T_{n,m} / \sqrt{2} $$

 Thus, we get
$$D \,\,\, = \,\,\, 2 \,  \left| O P^{\prime} \right|
\,\,\, \leq \,\,\, \sqrt{2} \, T_{n,m} $$

 It can be seen, for example from Fig. \ref{SingularNet}, 
that the given relation is also the exact upper bound for 
the value $\, D \, $.

\section{Conclusion}
\setcounter{equation}{0}

  In this paper, we consider ``scaling'' properties, namely, 
the parameters of the growth rate of level lines and the regions 
$\, V ({\bf r}) < \epsilon \, $ (or $\, V ({\bf r}) > \epsilon $) 
near the percolation threshold, for a special class of quasiperiodic 
potentials with eightfold rotational symmetry. In the study, 
we used an auxiliary ``extended'' family of quasiperiodic 
potentials, as well as a set of ``magic'' angles arising in 
this family. The study of the ``scaling'' properties of the 
potentials $\, V ({\bf r}) \, $ allows us to note some of their 
similarities and distinctive features in comparison with 
various models of random potentials on the plane. In fact, 
many similar properties are also possessed by quasiperiodic 
potentials of the ``extended'' family, which allows to consider
them also as an interesting model of random potentials with 
long-range order.

\end{document}